\documentclass[prl,aps,twocolumn,showpacs,superscriptaddress,nofootinbib]{revtex4}
\usepackage{graphicx} 
\usepackage{dcolumn}  
\usepackage{bm}       
\usepackage{amsmath}
\usepackage{epsfig}

\begin{document}
\title{Total light absorption in graphene}
\author{Sukosin Thongrattanasiri}
\affiliation{Instituto de \'Optica - CSIC, Serrano 121, 28006
Madrid, Spain}
\author{Frank H. L. Koppens}
\affiliation{ICFO-Institut de Ci\'encies Fot\'oniques, Mediterranean Technology Park, 08860 Castelldefels (Barcelona), Spain}
\author{F. Javier Garc\'{\i}a de Abajo}
\email[Corresponding author: ]{J.G.deAbajo@csic.es}
\affiliation{Instituto de \'Optica - CSIC, Serrano 121, 28006
Madrid, Spain}
\affiliation{Optoelectronics Research Centre, University of Southampton, Southampton SO17 1BJ, UK}


\begin{abstract}
We demonstrate that 100\% light absorption can take place in a single patterned sheet of doped graphene. General analysis shows that a planar array of small lossy particles exhibits full absorption under critical-coupling conditions provided the cross section of each individual particle is comparable to the area of the lattice unit-cell. Specifically, arrays of doped graphene nanodisks display full absorption when supported on a substrate under total internal reflection, and also when lying on a dielectric layer coating a metal. Our results are relevant for infrared light detectors and sources, which can be made tunable via electrostatic doping of graphene.
\end{abstract}
\maketitle



Beyond their fundamental interest in quantum theory and cosmology, blackbodies have practical relevance in light absorption and emission devices (e.g., detectors, photovoltaics, and optical sources). Inspired by nature, the following two routes have been investigated to achieve perfect absorption: (i) diffusion in disordered lossy surfaces, such as black silver, has led to engineered materials made of carbon nanotubes that exhibit extraordinary broadband light absorption \cite{MJK09}; (ii) alternatively, ordered periodic structures, which explain the {\it moth eye} effect observed in nocturnal insects \cite{CH1973}, have been pioneered by experimental and theoretical studies showing total light absorption (TLA) in the visible using metallic gratings \cite{HM1976,NMM1978,PMM08}. Similar phenomena have been reported at infrared (IR) \cite{PTM94} and microwave \cite{BFS05,LSM08} frequencies. Recently, omnidirectional TLA \cite{paper107} has been realized using periodic surfaces supporting localized plasmon excitations.


The availability of high-quality graphene as a stable material with extraordinary (opto-)electronic properties \cite{NGM04,NGM05,BSH10} makes a compelling case for exploring its ability to harvest light for potential application to optoelectronics, with the advantage of being optically tunable via electrostatic doping \cite{CPB11}. However, a single sheet of homogeneous graphene is poorly absorbing \cite{MSW08} (about 2.3\% absorption), so the challenge is to transform it into a perfect absorber, for which we can rely on its power to host extremely confined plasmons \cite{JBS09,graphene_first}.

In this Letter, we show that a single sheet of doped graphene, patterned into a periodic array of nanodisks, exhibits 100\% light absorption. We first discuss the extinction cross-section of graphene disks, which can exceed by over an order of magnitude their geometrical area \cite{graphene_first}. These disks therefore belong to the class of absorbing {\it particles} that can be arranged in periodic planar arrays such that their cross section exceeds than the area of the unit cell. We further assume here that all diffracted beams are evanescent. Under these conditions, we derive a universal maximum absorption for any thin layer separating two different media. In particular, TLA is predicted when the transmission channel is suppressed, for instance under total internal reflection (TIR) by illumination from a prism substrate, and also when the particles are supported on a dielectric film on top of a thick metal. We show that a patterned graphene sheet displays these effects. Full absorption by an atomically thin carbon film is thus possible using currently available technology.

\begin{figure*}
\begin{center}
\includegraphics[width=180mm,angle=0,clip]{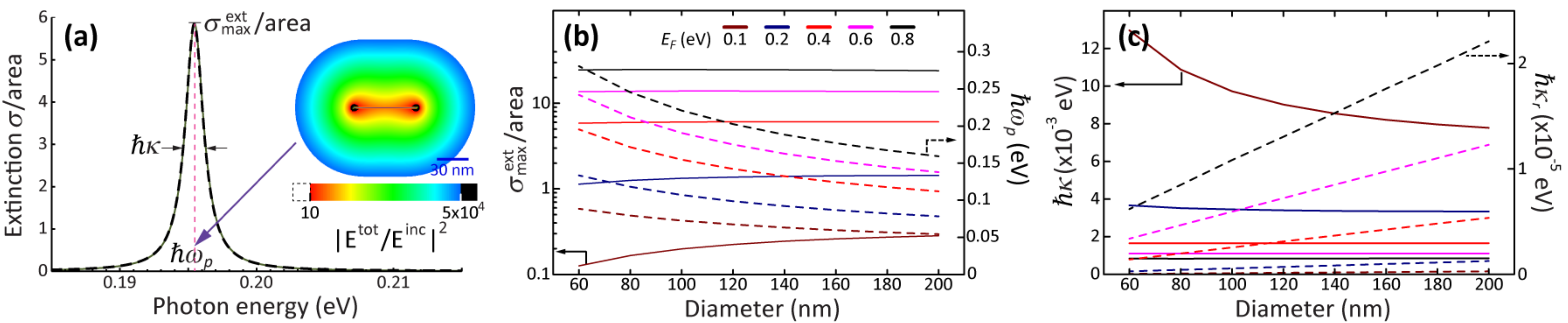}
\caption{Optical response of doped graphene nanodisks. {\bf (a)} Characteristic extinction cross-section of a nanodisk (60\,nm in diameter, and Fermi energy $E_F=0.4\,$eV), dominated by a pronounced dipolar plasmon and strong near-field enhancement. Solid curve: full electrodynamic calculation. Dashed curve: fit to a Lorentzian polarizability. Inset: near-field intensity relative to incident intensity for plane-wave illumination with polarization parallel to the disk. {\bf (b,c)} Size-dependence of the maximum extinction cross-section and the polarizability parameters [see Eq.\ (\ref{alpha})].} \label{Fig1}
\end{center}
\end{figure*}

{\it Optical response of graphene nanodisks.--} Doped graphene has been predicted to support strongly confined, long-lived plasmons \cite{JBS09}, which produce sharp IR resonances in nanodisks tens of times smaller than the light wavelength \cite{graphene_first}. Figure\ \ref{Fig1}(a) shows an example simulated with the boundary element method (BEM) \cite{paper040}. We then fit the polarizability of these small disks to \cite{VST96}
\begin{equation}
\alpha(\omega)=\frac{3c^3\kappa_r}{2\omega_p^2}\;\frac{1}{\omega_p^2-\omega^2-i\kappa\omega^3/\omega_p^2},
\label{alpha}
\end{equation}
where $\omega_p$ is the plasmon frequency, $\kappa$ is its decay rate, and $\kappa_r$ is the radiative contribution to $\kappa$. The extinction cross-section $\sigma^{\rm ext}=(4\pi\omega/c) {\rm Im}\{\alpha\}$ obtained from $\alpha$ [Fig.\ \ref{Fig1}(a), broken curve] is in good agreement with rigorous numerical solutions of Maxwell's equations (solid curve). We represent the parameters $\omega_p$, $\kappa$, and $\kappa_r$ in Fig.\ \ref{Fig1}(b-c) for different doping levels (quantified by the Fermi energy $E_F$) as a function of disk size. Here, we compute the conductivity of graphene in the local-RPA limit \cite{WSS06} with an intrinsic relaxation time $\tau=\mu E_F/ev_F^2$, where $v_F\approx c/300$ is the Fermi velocity and $\mu=10,000\,$cm$^2/$Vs is the measured DC mobility \cite{NGM04} (e.g., $\tau\approx10^{-13}\,$s for $E_F=0.1\,$eV).
Interestingly, the maximum extinction cross-section $\sigma^{\rm ext}_{\rm max}\approx(3\lambda^2/2\pi)\,\kappa_r/\kappa$ at the resonant wavelength $\lambda$ can exceed by over an order of magnitude the geometrical area, although the disks are far from perfect two-level systems (such as an atom without inelastic decay channels), for which $\sigma^{\rm ext}_{\rm max}\approx0.48\lambda^2$. This is a consequence of the fact that $\kappa\gg\kappa_r$.

\begin{figure*}
\begin{center}
\includegraphics[width=170mm,angle=0,clip]{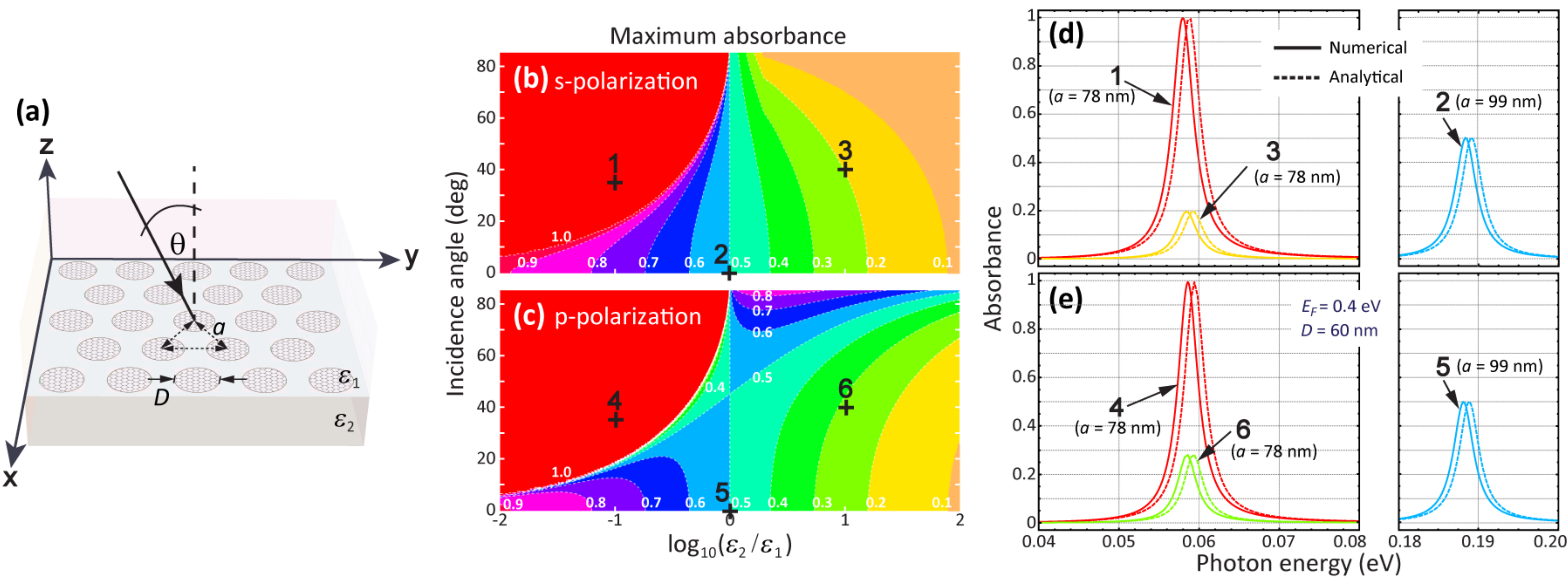}
\caption{Absorption by a layer of doped graphene nanodisks. {\bf (a)} Scheme of the geometry under consideration. {\bf (b,c)} Universal contours showing a maximum limit to the absorbance of a thin material layer sandwiched between two different media for $s$ and $p$ polarization as a function of incidence angle and dielectric contrast between both media. Light is incident from the $\epsilon_1$ medium. {\bf (d,e)} Realization of maximum absorption with arrays of graphene disks [see (a)] of diameter $D=60\,$nm, $E_F=0.4\,$eV, and different lattice spacings and angles of incidence (see labels). The lower-index medium has $\epsilon=1$ in all cases. The numerical labels in (d,e) correspond to the positions of the symbols in (b,c).} \label{Fig2}
\end{center}
\end{figure*}

{\it Universal limit to absorption by a thin layer.--} In a symmetric environment, the reflection coefficient $r$ of a thin material layer (much thinner than $\lambda$) determines the transmission coefficient $1\pm r$ (upper and lower signs are for $s$- and $p$-polarized light, respectively, as required to enforce the symmetry of the scattered field with respect to the layer plane; we use this convention from now on). Therefore, the maximum absorption ($1-|r|^2-|1\pm r|^2$) is limited to 50\% (corresponding to $r=\mp1/2$). This value can be increased by considering asymmetric environments, such as the patterned graphene sheet depicted in Fig.\ \ref{Fig2}(a).

A simple analysis leads to the maximum possible absorption produced by a thin, periodically structured material layer that is infinitesimally close to the interface between two media of different permittivities, $\epsilon_1$ and $\epsilon_2$. The reflection and transmission coefficients of this structure must read
\begin{equation}
R=r^0+(1\pm r^0)\eta, \;\;\; T=t^0\pm t^0\eta,
\label{eqr}
\end{equation}
where $r^0$ and $t^0$ are the Fresnel coefficients of the bare $\epsilon_1|\epsilon_2$ interface, while $\eta$ is the self-consistent amplitude of the wave scattered by the layer in response to both the external field and the multiple reflections within the layer-interface cavity. The scattered field is obviously symmetric for a layer of thickness $\ll\lambda$, and it contributes to Eq.\ (\ref{eqr}) with direct ($\eta$) and interface-reflected ($\pm r^0\eta$) components, as well as the corresponding transmission ($\pm t^0\eta$).
The absorbance $\mathcal{A}=1-|R|^2-{\rm Re}\{f\}|T|^2$, obtained from the integral of the Poynting vector over planes far from the structure, is thus an analytical function of the complex variable $\eta$, and its maximum admits the closed-form expressions
\begin{equation}
\mathcal{A}_{s,{\rm max}}=\frac{1}{1+{\rm Re}\{f\}}
\label{As}
\end{equation}
and
\begin{equation}
\mathcal{A}_{p,{\rm max}}=1-\frac{{\rm Re}\{f\}}{{\rm Re}\{f\}+(\epsilon_1/\epsilon_2)|f|^2}
\label{Ap}
\end{equation}
for $s$ and $p$ polarizations, respectively, where $f=(\epsilon_2/\epsilon_1-\sin^2\theta)^{1/2}/\cos\theta$ and $\theta$ is the angle of incidence from medium 1. The dependence of the universal maximum absorption on $\theta$ and $\epsilon_2/\epsilon_1$ predicted by Eqs.\ (\ref{As}) and (\ref{Ap}) is shown in Fig.\ \ref{Fig2}(b-c). Interestingly, it becomes 100\% under TIR from the high-index medium (i.e., for $\epsilon_1\sin^2\theta>\epsilon_2$, leading to ${\rm Re}\{f\}=0$), indicating that it is possible to simply suppress the reflection when transmission is already prevented by TIR.

{\it Absorption in particle and disk arrays.--} The universal maximum is just an upper limit to the absorption, but its derivation does not provide any clue on how to structure the material to reach such a limit, or whether it is at all possible to obtain it. However, a systematic inspection of graphene nanodisk arrays reveals that the universal limit is indeed reached by engineering the disk size, doping, and spacing. We show several examples in Fig.\ \ref{Fig2}(d-e), computed from a layer-KKR method \cite{SYM98}, in which the disks are described through their multipolar polarizability obtained from BEM (solid curves).

We can gain further insight into thin-layer absorption by describing each graphene nanodisk (or any other particle) through the polarizability [Eq.\ (\ref{alpha})]. For arrays of small period $a$ compared to the wavelength, the reflection coefficient of the layer reduces to \cite{paper090}
\begin{equation}
r=\frac{\pm iS}{\alpha^{-1}-G},
\label{r}
\end{equation}
where $G$ is a lattice sum whose imaginary part is exactly given by ${\rm Im}\{G\}=S-2(\omega/c)^3/3$, whereas its real part reduces to ${\rm Re}\{G\}\approx g/a^3$ for $\lambda\gg a$, with $g=5.52$ ($g=4.52$) for hexagonal (square) arrays. Here, $S=2\pi\omega/cA\cos\theta$ ($S=2\pi\omega\cos\theta/cA$) for $s$-polarized ($p$-polarized) light and $A$ is the unit-cell area. Equation\ (\ref{r}), together with Eq.\ (\ref{alpha}), permits obtaining the transmission and reflection coefficients of the combined array-interface system analytically (they are simply given by Eq.\ (\ref{eqr}) with $\eta=r(1\pm r^0)/(1-r^0r)$), leading to the dashed curves of Fig.\ \ref{Fig2}(d-e), in excellent agreement with full simulations.

In the symmetric configuration ($\epsilon_1=\epsilon_2$), combining Eqs.\ (\ref{alpha}) and (\ref{r}), and noticing that $\kappa\gg\kappa_r$, the maximum absorption (i.e., 50\%, corresponding to $r=\mp1/2$) is obtained at a frequency $\omega\approx\omega_p-3g\kappa_r/4(\omega a/c)^3$ (slightly redshifted with respect to the single-disk resonance $\omega_p$), under the condition $\zeta=1/2$, where we define the parameter
\begin{equation}
\zeta=\frac{A}{\sigma^{\rm ext}_{\rm max}}\times
\begin{cases}
\cos\theta, & s\;{\rm polarization}, \\
\cos^{-1}\theta, & p\;{\rm polarization}.
\end{cases}\nonumber
\end{equation}
This condition can be fulfilled by adjusting the spacing between disks, provided $\sigma^{\rm ext}_{\rm max}$ is sufficiently large compared to the disk area. For asymmetric environments ($\epsilon_1\neq\epsilon_2$), the analysis becomes more involved, but similar redshifts are obtained for large enough $\sigma^{\rm ext}_{\rm max}$.

\begin{figure}
\begin{center}
\includegraphics[width=70mm,angle=0,clip]{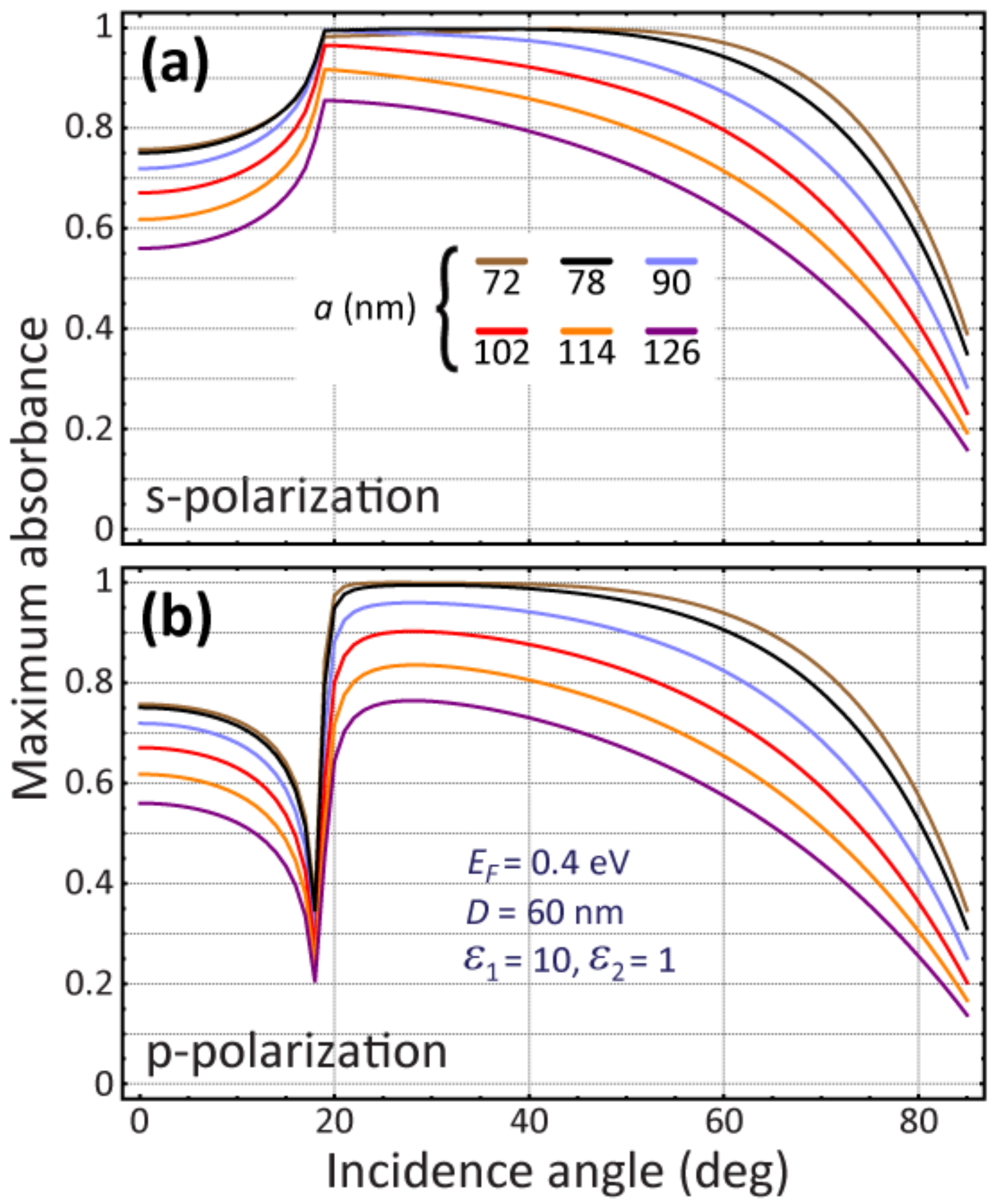}
\caption{Angular dependence of the peak absorbance for $s$ (a) and $p$ (b) polarization in arrays of graphene nanodisks [see Fig.\ \ref{Fig2}(a)]. Each curve represents a fixed sample with parameters as shown by labels.} \label{Fig3}
\end{center}
\end{figure}

The robustness of TLA under TIR is explored in Fig.\ \ref{Fig3}, where the peak absorbance is represented as a function of incidence angle for various values of the lattice spacing $a$, using the same disks in all cases [see parameters in Fig.\ \ref{Fig3}(b)]. Nearly TLA is observed within the $20^\circ$-$45^\circ$ range of incidence angles under critical coupling conditions \cite{paper107}.

\begin{figure*}
\begin{center}
\includegraphics[width=170mm,angle=0,clip]{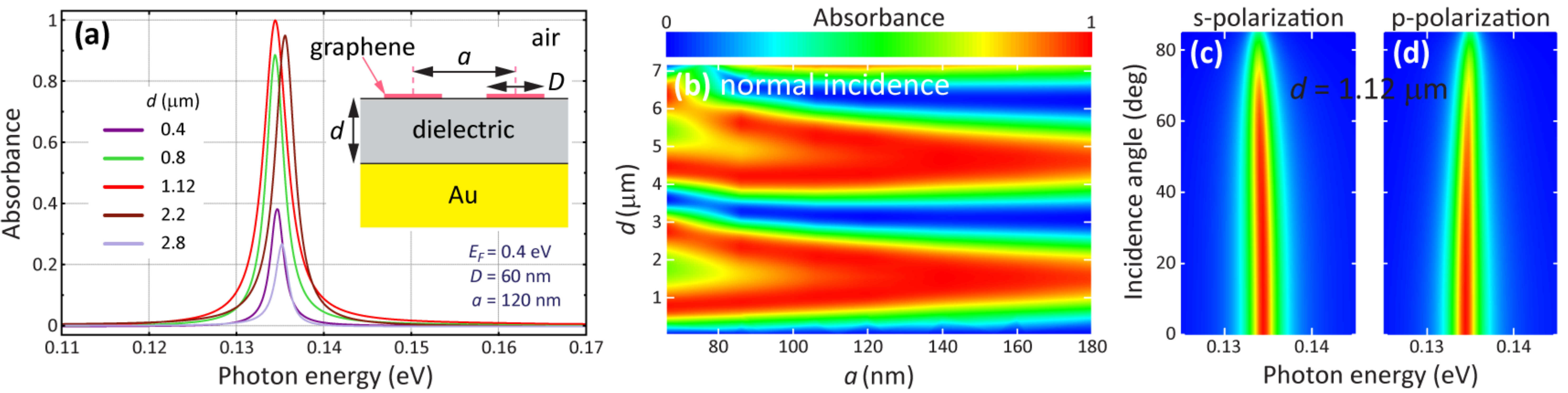}
\caption{Total absorption with graphene on a coated metallic substrate. {\bf (a)} Normal-incidence absorbance by graphene disk arrays supported on a dielectric-coated gold surface (see inset) for various values of the dielectric film thickness ($\epsilon_{\rm dielectric}=2.1$, $\epsilon_{\rm Au}$ from \cite{JC1972}). {\bf (b)} Peak absorbance as a function of dielectric thickness $d$ and lattice spacing $a$ for the same disks as in (a). {\bf (c,d)} Incidence-angle and photon-energy dependence of the absorbance under the conditions of (a) for $d=1.12\,\mu$m.} \label{Fig4}
\end{center}
\end{figure*}

{\it Omnidirectional total light absorption.--} The above prediction of TLA under TIR can be generalized to systems in which the transmission channel is suppressed. In particular, we consider a graphene array placed above a metallic substrate. The graphene plasmons disappear close to a metal, but this problem is solved by introducing an intermediate dielectric coating layer [see inset of Fig.\ \ref{Fig4}(a)]. In this configuration, the metal can naturally act as a backgate element to dope the graphene. It is easy to see that reflection is canceled when the substrate (i.e., metal plus coating) reflection coefficient is $r^0=-r/(1\pm2r)$. The coating-layer thickness $d$ can be adjusted to match the required phase in this expression, and the condition is simply stated as $1/|r^0|=|2\pm1/r|\ge1$, where the equality applies to non-absorbing substrates (e.g., a dielectric under TIR, or noble metals in the IR). Figure\ \ref{Fig4}(b) shows TLA under this scheme for properly adjusted values of $a$ and $d$. Obviously, the phase-matching condition is periodically satisfied along the $d$ axis. Interestingly, the incidence-angle and polarization dependence of the absorption is very weak [Fig.\ \ref{Fig4}(c-d)], pointing at the possibility of omnidirectional absorption. To understand this, we combine Eqs.\ (\ref{alpha}) and (\ref{r}) together with the relation $|2\pm1/r|=1$ (i.e., the condition of no reflection for $|r^0|=1$), from which we find two frequencies for TLA, $\omega=\omega_p-3g\kappa_r/4(\omega a/c)^3\pm(\kappa/2)\sqrt{1/\zeta-1}$, under the condition $\zeta<1$; only the right-most term depends on polarization and angle of incidence (through $\zeta$), so total absorption can be made omnidirectional when the second term in the resonant $\omega$ dominates (i.e., for $(\omega a/c)^3\ll\kappa_r/\kappa$), provided $\sigma^{\rm ext}_{\rm max}$ is of the order of $A$.

{\it Concluding remarks.--} Planar textured materials can produce large light absorption, which reaches 100\% when the transmission channel is suppressed. We predict that this effect takes place using a single sheet of doped patterned graphene either under TIR or when the carbon layer is deposited on dielectric-coated metal. Full absorption in graphene opens interesting possibilities for light detection (e.g., via electron-hole separation \cite{LKF11}, thermoelectric reading \cite{SRM11}, etc.). This type of detector should inherit the electrical tunability of graphene, which can be used for direct, efficient spectral analysis of IR light with relative spectral resolution $\sim\kappa_r/\kappa$. Additionally, thermally heated samples can be useful as narrow-band IR sources in virtue of Kirchhoff's law. We further predict perfect absorption in planar particle arrays under the condition that the absorption cross-section of an individual particle is comparable to the area of the unit cell. This condition is actually fulfilled by many non-ideal two-level atoms/molecules (e.g., NV centers). Our results could be tested with optically trapped atomic arrays, although graphene nanodisks are more convenient because they provide a practical implementation that is stable under ambient conditions. Noble-metal nanoparticle arrays can also exhibit similar effects in the near IR, with their spacing controlled through dielectric coating \cite{ULM01}.

This work has been supported by the Spanish MICINN (MAT2010-14885 and Consolider NanoLight.es), Fundaci\'o Cellex Barcelona, and the European Commission (FP7-ICT-2009-4-248909-LIMA and FP7-ICT-2009-4-248855-N4E).


\end{document}